# Vibrational Spectroscopic Detection of a Single Virus by Interferometric Mid-Infrared Photothermal Microscopy

Yi Zhang[1]#, Celalettin Yurdakul[2]#, Alexander J. Devaux[3], John H. Connor [3], M. Selim Ünlü*[1,2,4], and Ji-Xin Cheng*[1,2,4]

[1]Department of Physics, Boston University, Boston, MA 02215, USA
[2]Department of Electrical and Computer Engineering, Boston University, Boston, MA 02215, USA
[3]Department of Microbiology and National Infectious Diseases Laboratories, Boston University School of Medicine, Boston, Massachusetts 02118, USA
[4]Department of Biomedical Engineering, Boston University, Boston, MA 02215, USA

*Corresponding authors. E-mail: jxcheng@bu.edu, selim@bu.edu. #equal contributions.

**ABSTRACT:** We report a confocal interferometric mid-infrared photothermal (MIP) microscope for ultra-sensitive and spatially resolved chemical imaging of individual viruses. The interferometric scattering principle is applied to detect the very weak photothermal signal induced by infrared absorption of chemical bonds. Spectroscopic MIP detection of single vesicular stomatitis viruses (VSVs) and poxviruses is demonstrated. The single virus spectra show high consistency within the same virus type. The dominant spectral peaks are contributed by the amide I and amide II vibrations attributed to the viral proteins. The ratio of these two peaks is significantly different between VSVs and poxviruses, highlighting the potential of using interferometric MIP microscopy for label-free differentiation of viral particles. This all-optical chemical imaging method opens a new way for spectroscopic detection of biological nanoparticles in a label-free manner and may facilitate in predicting and controlling outbreaks of emerging virus strains.

Viruses represent both a vexing public health problem and an increasingly common source of therapeutics and vaccines. Viruses are responsible for numerous epidemic outbreaks, including the 1918 influenza pandemic[1], Zika[2], Ebola[3], and the ongoing COVID-19[4]. Attenuated viruses are the basis for more than a dozen vaccines[5]. Genetically engineered viruses are approved for gene therapy[6], direct oncolytic cancer therapy[7], and CAR-T cell treatments[8]. The critical unit in both harmful and helpful virus functions is the virion, an extensively loaded nanoscale particle that shields and then delivers the infectious viral unit into susceptible cells. The make-up and function of individual virions is an evolving area of study, benefitting from powerful techniques such as electron microscopy, atomic force microscopy[9], reflectance imaging microscopy[10], and structured illumination microscopy using fluorescently tagged virions[11]. Each of these approaches has advantages, but none to date have shown the ability to directly probe the chemical content or signature of an individual virion on a single particle scale.

Interferometric biosensing and microscopy have been extensively studied for sensitive, high-resolution, and ultrafast nanoscale specimen studies.[12-15] The interferometric imaging enhances the weakly scattered light from a nanoparticle of interest via interference with a stronger reference field that brings the scattered signal into the shot-noise limit of the imaging system, which otherwise falls short of the detector noise in dark-field detection as a consequence of drastic signal drop[16-18]. In wide-field modality using a layered substrate, interferometric reflectance imaging sensor (IRIS) has been demonstrated to detect viruses[19] and exosomes[20] without labels as well as single protein[21] and nucleic acid[22] molecules by labeling with plasmonic nanoparticles. Interferometric microscopy pushes the sensitivity beyond the bright-field optical microscopy down to a single protein in both wide-field[23, 24], and confocal scanning modes[25, 26] using a glass substrate. To date, interferometric studies of biological nanoparticles have been limited to affinity-specific molecular information, whereas the chemical composition of particles has remained inaccessible.

To obtain molecular information beyond the surface affinity, Vibrational spectroscopy based on either Raman scattering or infrared absorption is commonly employed in biological studies. These methods identify the chemical content of a substance based on molecular vibrational fingerprints. For Raman scattering, the cross-section is extremely small – approximately 1 out of $10^{10}$ incident photons are emitted through Raman scattering. Surface-enhanced Raman scattering (SERS) has shown great potential for single molecule detection at a plasmatic hot spot[27-30]. The detection reliability of SERS is, however, relied on the quality of the substrate. Besides, because the hot spot (< 10 nm) is smaller than a virus (~ 100 nm), SERS is not able to probe the entire content of a virus. Tip-enhanced Raman spectroscopy

(TERS) detection of the virus is reported[31], where the reliability strongly depends on the tip quality. Recently developed coherent Raman scattering microscopy allowed high-speed vibrational imaging of cells and tissues[32, 33]. Yet, coherent Raman scattering imaging of a single virus has not been demonstrated, partly due to the small cross-section of Raman scattering from a nanoscale viral particle.

In comparison to inelastic Raman scattering, infrared absorption has 6-8 orders of magnitude larger cross section[34]. The infrared spectra deliver information on proteins as well as on interactions between protein subunits and nucleic acids[35, 36]. Yet, spectroscopic detection of a single virus by conventional FTIR is hampered by the intrinsically low spatial resolution on the micron scale[37]. AFM-IR, which offers nanoscale spatial resolution and high detection sensitivity, is capable of providing a single virus infrared spectrum[38]. Yet, AFM-IR requires sample contact, thus suffering sample damage risk. Thus, a contact-free, easy to operate, and highly sensitive method for single virus detection is desired.

Mid-infrared photothermal (MIP) imaging is an emerging technique[39-43] meeting these requirements. In MIP microscopy, a visible beam is deployed to sense the photothermal effect induced by infrared absorption of molecules, providing sub-micron spatial resolution defined by the visible probe beam. MIP signals can be detected in either scanning mode [39-43] or widefield mode[44-47]. On the excitation side, MIP can be implemented with an IR/Visible co-propagating or counter-propagating geometry. In the latter, an objective lens of a high numerical aperture can be used to focus the visible beam and a spatial resolution of 300 nm has been reached[41]. More recently, researchers at the University of Notre Dame pushed the detection limit to 60 nm polymer particles with a balanced detection scheme[42].

Here, we harness the interferometric scattering principle in confocal configuration to push the sensitivity limit of MIP microscopy to an unprecedented level towards single virus detection. We report, for the first time, vibrational fingerprint spectra of single poxvirus and vesicular stomatitis virus (VSV) recorded with a counter-propagating interferometric MIP microscope.

**Figure 1. Illustration and principles of interferometric MIP microscopy**. (a) Schematic. A pulsed mid-infrared pump beam is provided by a quantum cascaded laser (QCL) laser and focused into a sample by a reflective objective (RO). A continuous visible (green) probe beam is focused on the sample by a water immersion objective lens (OL). To measure the IR power, a small fraction of the IR beam is reflected by a $CaF_2$ glass onto a mercury cadmium telluride (MCT) detector. The interferometric signal ($E_{det}$) is collected in an epi-illumination configuration using a beam splitter (BS). A silicon photodiode (PD) detects $E_{det}$ after filtering by a pinhole (PH) in a confocal configuration. The 4f system consisting of two achromatic doublets L1 and L2 provides access to the confocal conjugate plane for the filtering. (b) Illustration of the counter propagation interferometric scattering detection principle. (c) Electronic connections for system control and signal detection. PD is connected to a resonant amplifier (RA) to amplify the detected probe beam. Lock-in amplifier (LIA) demodulates the PD signal to isolate the photothermal signal. PC controls the motorized scanning stage and data acquisition. The IR intensity detected by MCT is also received by LIA for normalization of the photothermal signal by IR power at each wavenumber. The lock-in sends a reference signal to trigger the laser at 100 kHz frequency.

## Methods

### Experimental setup

A schematic is shown in Figure 1. A pulsed mid-IR pump beam, generated by a tunable (from 1000 to 1886 cm$^{-1}$) quantum cascade laser (QCL, Daylight Solutions, MIRcat-2400) operating at 100 kHz repetition rate, passes through a calcium fluoride ($CaF_2$) cover glass and then is focused onto the sample through a gold-coating reflective objective lens (52×; NA, 0.65; Edmund Optics, #66589). A continuous-wave probe laser (Cobolt, Samba 532 nm) beam is focused onto the same spot from the opposite side by a high NA refractive objective (60×; NA, 1.2; water immersion; Olympus, UPlanSApo). The probe beam is aligned to be collinear to the mid-IR pump beam to ensure the overlap of the two foci to achieve a good signal level. A scanning piezo stage (Mad City Labs, Nano-Bio 2200) with a maximum scanning speed of 200 μs/pixel is used to scan the sample. Before the photodiode, a confocal pinhole is placed to pass only the scattered photons from the nanoparticles and the reflected photons from the top surface of $CaF_2$ glass while blocking the photons reflected from other surfaces. In the imaging procedure, the nanoparticles spotted on a $CaF_2$ cover glass are first localized by the backward interferometric signal. The sample stage is adjusted axially to maximize the interferometric contrast. Then, the pulsed infrared pump beam illuminates the sample. The modulated scattering is collected by a photodiode and the MIP signal is extracted by a lock-in amplifier. Before the reflective objective, the infrared laser passes through a $CaF_2$ coverslip and the refection of the infrared laser is measured by a mercury cadmium telluride (MCT) detector for normalization of IR power at each wavelength. A laboratory-built resonant circuit, with its resonant frequency (103.8 kHz, gain 100) tuned to the repetition rate of the QCL laser, is used to amplify the photocurrent from the photodiode before it is sent to the lock-in amplifier (Zurich Instruments, HF2LI) for phase-sensitive detection of the MIP signal.

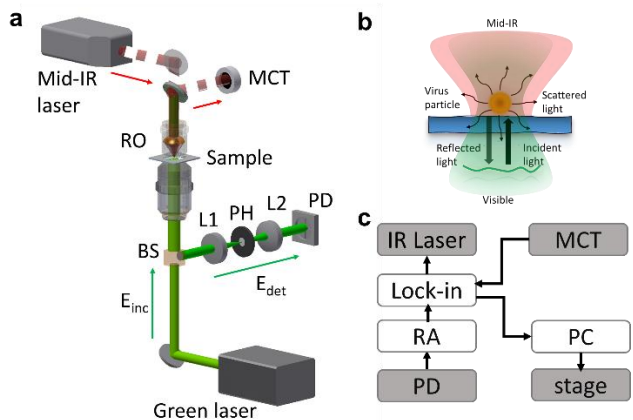

To normalize the MIP signal by IR power at each wavenumber, the mid-IR intensity detected at the MCT is sent to another lock-in input channel. A computer is used to synchronize the QCL wavelength tuning, stage scanning, and data acquisition. The photothermal spectra are processed using in-house scripts written on MATLAB. The raw photothermal signal is normalized by the mid-infrared laser power spectrum collected by the MCT detector at a step size of 5 cm$^{-1}$. For detection of the virus, the photothermal spectra are averaged 5 times to achieve a high signal to noise ratio (SNR).

**Principle of interferometric photothermal signal detection**

Our system's detection mechanism relies on the ultra-sensitive interferometric scattering imaging principle[18] which employs a common-path interferometry. This configuration enables highly sensitive and stable interferometric imaging of nanoparticles captured on the substrate surface. The light impinging on particle scatters and reflects off the glass/air interface. The complex valued reflected field $\boldsymbol{E_r} = E_r e^{j\phi_r}$ and scattered field $\boldsymbol{E_s} = E_s e^{j\phi_s}$ interfere at the detector: $I_{det} = |\boldsymbol{E_r} + \boldsymbol{E_s}|^2$. This recorded interference signal can be formulated as follows:

$$I_{det} = E_r^2 + E_s^2 + 2E_r E_s \cos(\theta) \quad (1)$$

The first term denotes the reference field intensity $I_r = |E_r|^2$, the second term denotes the scattering intensity $I_s = |E_s|^2$, and the third term denotes the real valued interferometric cross term $2Re(\boldsymbol{E_r E_s^*})$. The phase term $\theta = \phi_r - \phi_s$ is the phase difference between the reference and scattered fields. The strength of the scattered field amplitude scales with the sample's polarizability α. In the dipole limit where particle radius $r$ is much smaller than illumination wavelength $r \ll \lambda$, the polarizability for a nanosphere[48] is given by:

$$E_s \propto \alpha = 4\pi\epsilon_m r^3 \frac{\epsilon_p - \epsilon_m}{\epsilon_p + 2\epsilon_m} \quad (2)$$

where $\epsilon_p$ and $\epsilon_m$ respectively denote the dielectric permittivity of particle and the surrounding medium. Considering small nanoparticles, the scattering intensity term which has $r^6$ dependence generates signal levels well below the shot-noise limited detection and quickly vanishes against the reference field intensity. Therefore, the scattering intensity term in eq. 1 becomes negligible due to the sample's weakly scattering nature. The resulting signal constitutes the background signal ($I_r$) and the interferometric cross term. The interferometric term realizes the linear detection of the weak scattered field with a strong reference field enhancement. Moreover, such detection allows for the detector to operate at the shot-noise limited performance. After background subtraction, the interferometric signal can be expressed as follows:

$$S = I_{det} - I_r = 2E_s E_r \cos(\theta) \quad (3)$$

The phase difference contains the accumulated Gouy phase, field propagation phase, and the sample's internal field phase. The sinusoidal term modulates the signal contrast depending on the axial position with respect to the objective focus. This term can be optimized to maximize the contrast by defocusing[49].

The IR pulse vibrationally excites the molecules inside a sample. This heats the sample depending on the optical absorption cross-section and IR-pulse intensity. The absorption then causes a local temperature rise of $\Delta T$ at the particle's vicinity. This photothermal effect induces a subtle change in the particle's refractive index and size due to thermal expansion. As seen from eq. 2, this effect modifies the nanoparticle's polarizability and thus the scattered field amplitude. The induced signal change can be written as:

$$\Delta S = 2E_r \Delta E_s \quad (4)$$

where $\Delta E_s = E_s(T_0 + \Delta T) - E_s(T_0)$ is the difference of the scattered field amplitudes with a pre-IR pulse temperature $T_0$. We assume here the particle as a uniform heat source which is a valid assumption for very small nanoparticles compared with the micron-scale IR wavelength. Similar to the DC case, the interferometric photothermal imaging realizes linear detection of the change in the scattering amplitude. The photothermal signal can be measured as the intensity modulation at the detector. One can approximate the modulation depth $\Delta S/S$ by the means of the particle's polarizability change as follows:

$$\frac{\Delta S}{S} \approx \frac{\Delta \alpha}{\alpha} \approx 3\Delta T \left( \alpha_r + \frac{2\epsilon_p}{(\epsilon_p + 2\epsilon_m)(\epsilon_p - \epsilon_m)} \alpha_n \right) \quad (5)$$

where $\alpha_r = \frac{1}{r} dr/dT$ and $\alpha_n = \frac{1}{n} dn/dT$ are the thermal expansion and thermo-optic coefficients, respectively. This approximation holds for $\Delta r \ll r$ and $\Delta n \ll n$. The modulation fraction becomes in the orders of 0.01% per 1 K for PMMA beads for given $dr/dT = 90\times10^{-6}$ mK$^{-1}$ and $dn/dT = -1.1\times10^{-4}$ K$^{-1}$ [43]. The lock-in amplifier is utilized to extract this very minute signal.

**Virus sample preparation**

The recombinant vesicular stomatitis virus expressing green fluorescent protein envelope (rVSV-G-eGFP) was generated as described earlier [50] and the recombinant vaccinia virus expressing Venus fluorescent protein (rVACV-A4L-Venus) was generated as described earlier [51]. To load viruses onto the substrate surface, 100 µL of either VSV or poxvirus stock was incubated on a CaF2 coverslip for 1 hour at room temperature. The rVSV and rVACV concentrations were ~3×10$^8$ PFU/mL and ~2×10$^8$ PFU/mL, respectively. Following incubation, coverslips were examined with fluorescence microscopy to ensure that virus was present. All virions were crosslinked and inactivated using 1.0 mL 4% formaldehyde for 1 hour and then dried.

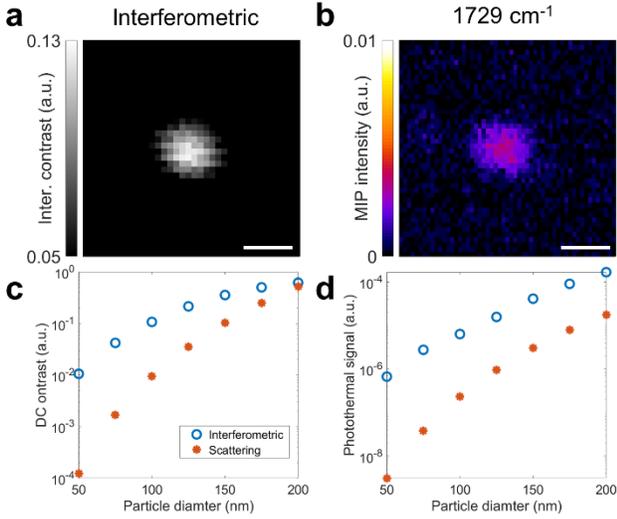

**Figure 2. Experimental demonstration and theoretical calculations of interferometric detection.** (a) Interferometric scattering image of 100 nm PMMA bead, (b) Interferometric MIP image of the 100 nm PMMA bead at 1729 cm$^{-1}$. Probe power on the sample: 40 mW; IR pump power on the sample: 8.5 mW. Scale bars: 500 nm; pixel dwell time: 20 ms; image acquisition time: 19.6 s. (c) Contrast comparison between interferometric and scattering signals for PMMA bead with particle size range from 50 nm to 200 nm in diameter. (d) Photothermal signal corresponds to the signals in **a**. Temperature rise is 1 K.

## Results and Discussion

### Experimental detection of 100 nm particles and theoretical calculations

To demonstrate the capability of interferometric photothermal detection, we first image 100 nm poly (methyl methacrylate) (PMMA) beads. These PMMA beads present a suitable model for system characterizations since their size resembles the viral particles in our experiments and importantly PMMA has a dielectric constant (n≈1.5) similar to that of viruses. Figure 2a shows a high SNR interferometric scattering image with 0.13 particle contrast. The interferometric contrast is defined as normalized, background-subtracted signal [18]. This indicates that scattering intensity is approximately 400 times smaller than the background (reflected field) intensity. The interferometric detection reveals highly sensitive detection of very weak signals from small particles which may be indistinguishable from the background in the conventional light microscopy. This comes from the fact that residual reflection (~ 4%) from the substrate and particle medium interface. Compared to forward detection schemes where most of the incident light reaches the detector, the reflection geometry in interferometric detection of small particles can improve the contrast by approximately 25 times. We note this is valid in the dipole approximation. Since large particles scatter mostly in the forward direction, we expect to achieve similar contrast levels in both reflection and forward systems.

The nanoscale particles can be modeled in the dipole limit as discussed in the theory section. We perform the theoretical simulations on a custom-developed electromagnetic simulation software on Matlab [52]. This dipole model shows a great agreement with experiments and a more comprehensive electromagnetic model for particles up to 200 nm [53]. We extend this simulation to interferometric imaging configuration. Our simulation can be compartmentalized into two main parts: (1) substrate and particle geometry and dielectric coefficients to define the dipole moment and (2) optical system parameters including objective numerical aperture and illumination function to calculate the point spread functions. Accordingly, we assume the particles are placed on top of a CaF$_2$ substrate with a height of particle's radius (r). The dipole moment is calculated using polarizability tensor and total driving electric field ($E_{inc}$). The scattered fields are then calculated using far-field Green's functions. The reflected field is calculated using Fresnel's coefficients considering incidence angle and polarization. Both reflected fields and scattered far fields are mapped into the image space using angular spectrum representation integral. Then, we coherently sum the fields using Eq. 1. Accordingly, we calculate the field contributions in the detected signals for a size range from 50 to 200 nm. The theoretical interferometric contrast for 100 nm PMMA bead is 0.11 which shows an agreement with the experimental value. The incident field amplitude remains constant in all cases. As shown in Figure 2c, the interferometric signal becomes more dominant compared to scattering intensity as the particle size decreases. This comes from the fact that the interferometric detection has r$^3$ dependency compared with the r$^6$ dependency of the scattering intensity. Therefore, the scattering intensity only detection schemes for nanoscale particles drastically drops below the detector's shot noise limit.

We next characterize the MIP image of the 100 nm PMMA bead as shown in Figure 2b. The MIP signal has ~13 SNR which is close to the sensitivity limit of our system. The modulation depth is 0.012 %, indicating that photothermal induced temperature increase in the bead is around 1 to 2 K. Such temperature change is significantly lower than the previously reported studies using optical parametric oscillator laser[42] which has shorter pulse width and relatively larger pulse energy. The temperature rise and decay time for sub-200 nm particles are on the order of the tens of nanoseconds which is much shorter than the QCL laser pulse width (1 μs). Therefore, a lower temperature change is expected in our experiments. To improve sensitivity, one can employ short pulse laser sources with high pulse energy. We further simulate the photothermal signal corresponding to the interferometric scattering contrasts in Figure 4c. In the calculations, we update the particle size and refractive index by assuming a 1 K temperature rise. Figure 4d shows the absolute value of the signal change induced by the photothermal effect. The results indicate that interferometric detection generates ~25 times more signal for a 100 nm PMMA bead. Thus, interferometric detection of the photothermal effect has a significant advantage over the scattered intensity only detection.

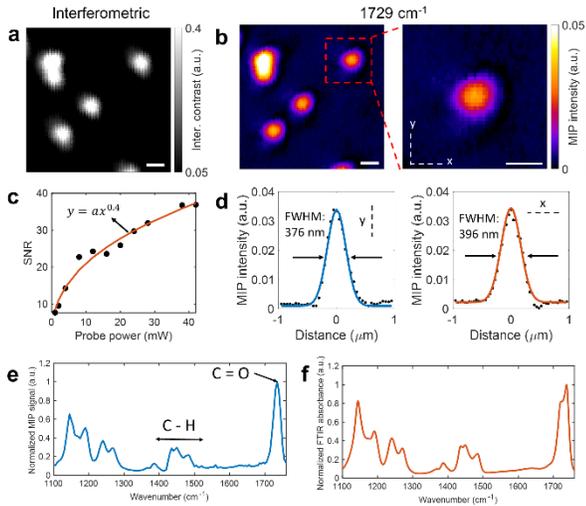

**Figure 3. Performance of interferometric MIP microscopy.** (a) Interferometric scattering image of 200 nm PMMA beads, (b) (Left) MIP image in **a** and (Right) zoom-in region of a single PMMA bead. Probe power on the sample: 30 mW; IR pump power on the sample: 5 mW at 1729 cm$^{-1}$. Scale bars: 500 nm; pixel dwell time: 20 ms; image acquisition time: 127 s. (c) SNR characterization of MIP signal showing that shot-noise limit is reached. (d) (Left) Vertical and (Right) horizontal cross-section profiles across the enlarged single bead in **b**. The Gaussian fitted full width at half maximums (FWHM) are respectively 376 nm and 396 nm. (e) MIP spectrum obtained from a single PMMA bead and (f) FTIR acquired from a PMMA film. MIP acquisition time: 10 ms per wavenumber.

### Sensitivity, resolution, and spectral fidelity

We then evaluate the sensitivity and resolution of the MIP system by measuring 200 nm diameter since they generate a higher SNR signal compared to 100 nm beads. This is particularly important for spectral characterization to capture low absorption bonds with high data fidelity. Figure 3a shows an interferometric scattering image of PMMA beads. We employ confocal detection to minimize the unwanted reflection from the other surfaces in the optical path. This improves the SNR by suppressing the noise from other sources contributing to the shot-noise. To obtain high quality images, we extended the dwell time of each pixel to 20 ms and applied a small pixel step size of 50 nm. With these improvements, we obtain an SNR of 96 in Figure 3b. The MIP images are taken by tuning the IR laser to 1730 cm$^{-1}$ which is the resonance peak corresponds to acrylate carboxyl (C=O stretching) group absorption in PMMA. To evaluate the spatial resolution of the MIP signal, we measured the intensity profile across the single bead in Figure 3b. The measured full width at half maximum (FWHM) along the vertical and horizontal axes are 376 nm and 396 nm, respectively. The slight deviation along the x and y axes could be attributed to the scanning stage's mechanical stability. After deconvolution by particle diameter, we obtain a spatial resolution of 319 nm. This value is consistent with the previous studies using the same counter-beam geometry[41]. Furthermore, we characterize the system's shot noise performance by acquiring MIP images at varying probe powers. The exponential fit coefficient is 0.4 which is close to the theoretical value of 0.5. We then measure single particle spectra spanning from 1100 cm$^{-1}$ to 1750 cm$^{-1}$ (Figure 3e). Both resonance peaks of C=O and C-H stretching bonds show the distinguished spectral signature in the MIP system. The spectral fidelity is confirmed by comparing the MIP spectral profile to the reference spectrum of PMMA collected by an FTIR spectrometer (Figure 3f).

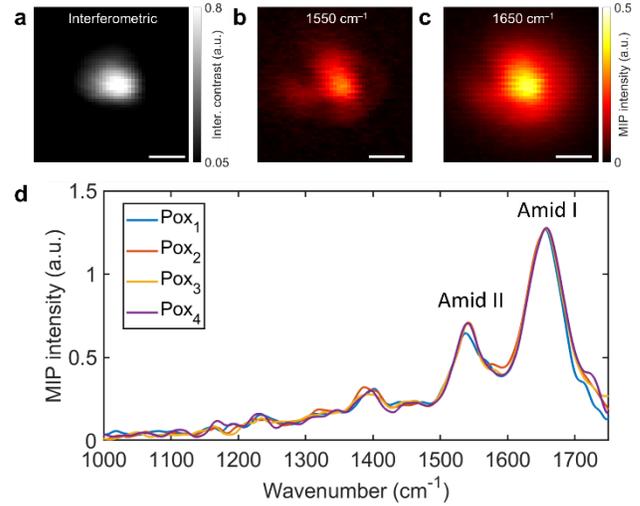

**Figure 4. Interferometric MIP imaging and spectroscopy of single poxviruses.** (a) Interferometric scattering image of a poxvirus. Probe power on the sample: 30 mW. (b) MIP image of the same poxvirus at the 1550 cm$^{-1}$. IR pump power on the sample: 18 mW. (c) MIP image of the same virus at the 1650 cm$^{-1}$. IR power: 20 mW; Scale bars: 500 nm; pixel dwell time: 20 ms; image acquisition time: 46.4 s. (d) MIP spectra of four randomly chosen poxviruses.

### Fingerprinting individual viruses

Based on the above characterization, we explore the potential of MIP microscopy for single virus detection. We first demonstrate our system's capability on poxviruses. Poxvirus belongs to a family of double stranded DNA viruses[54]. They are brick-shaped (240 nm by 300 nm) with a well-packed internal structure. Since the lateral resolution of the imaging system is insufficient to resolve the shape and morphology of the virus, the viruses appear as diffraction-limited spots. Figure 4a shows the interferometric scattering image of a poxvirus, where the scattering from the poxvirus is accompanied by a large background contributed by reflected photons. Importantly, the interference between the reflected and scattered light enhances virus contrast which becomes detectable in the shot-noise-limited regime. Figure 4b and Figure 4c shows the MIP image of the same virus in Figure 4a, with the IR laser tuned to 1650 cm$^{-1}$ for the amide I band and 1550 cm$^{-1}$ for the amide II band of viral proteins. Both images show high contrast because the background is removed through lock-in detection. In the interferometric scattering image of the poxvirus, the scattered photons after interference with the reflected photons generated a 0.47 V signal on the photodiode and the background gave around a 0.87 V signal. Thus, we achieved around 80 % contrast in the interferometric image, which allowed us to detect the virus particle. For the MIP signal, according to the input range, output signal level, and internal amplification factor from the lock-in amplifier, the modulation depth is calculated to be around 0.1 %. Such changes can be readily extracted by a lock-in amplifier, which allows background-free chemical imaging of the virus. We then took MIP spectra in the fingerprint window (1000 to 1750 cm$^{-1}$). This procedure is repeated on four randomly chosen poxvirus. The spectra demonstrate high consistency, as shown in Figure 4d. We observe two dominated bands contributed by amide I and amide II vibrations, but a negligible signal from the phosphate vibration at 1080 cm$^{-1}$. The virus has a well-organized structure in which a protein shell covers the DNA inside. It is possible that vibrational absorption of DNA might not cause sufficient change in size or refractive index for MIP detection.

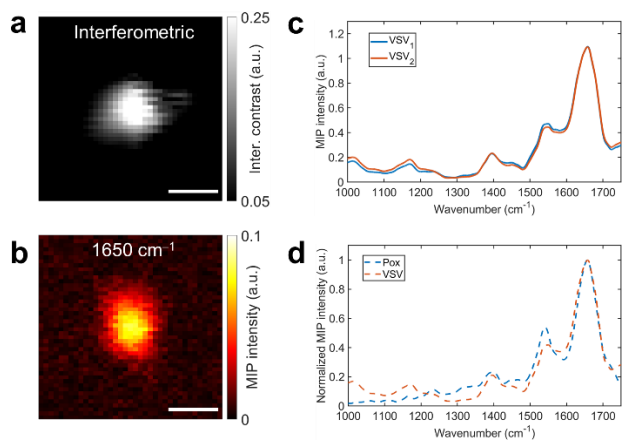

**Figure 5. Interferometric MIP imaging and spectroscopy of a single VSV.** (a) Interferometric scattering image of a single VSV and MIP signal with the IR laser tuned to 1650 cm$^{-1}$ (Amide I bond). Probe power on the sample: 30 mW; IR pump power on the sample: 20 mW. Scale bars: 500 nm; pixel dwell time: 20 ms; image acquisition time: 34 s. (c) MIP spectra of two VSV particles. (d) Each plot represents the averaged spectra of individual viruses in Figures 4d and 5c. To emphasize the spectral differences, each spectrum is normalized by the maximum MIP intensity at the amid I band.

Next, to test whether our technology can distinguish different virus types, we perform MIP detection of VSV. Figure 5a shows the interferometric scattering image of VSV. To obtain the MIP signal shown in Figure 5b, the IR laser is tuned to 1650 cm$^{-1}$ (Amide I band). A high SNR of 35 is obtained in the MIP image. Then, we acquire MIP spectra from the two randomly chosen individual viruses. As shown in Figure 5c, the spectral profiles are highly consistent in both viruses. More importantly, we observe significant spectrum differences between poxvirus and VSV. The amide I and amide II peaks are the major feature of the virus MIP spectrum. The amide I to amide II intensity ratio for poxvirus is found to be 0.51 and the ratio for the VSV is 0.42. Also, the gap between the amide I and amide II peaks are less obvious for VSV. Figure 5d further emphasizes the spectrum differences by comparing them side by side. These spectroscopic features are likely a reflection of the unit structure and molecular content of each virus and can be used to differentiate various types of viruses. As an indirect spectroscopic technique, SERS reporters have been successfully demonstrated in triplex assays for Ebola, malaria, and Lassa in which each specific reporter has distinguished Raman spectrum[55]. Here, our findings suggest that a direct multiplex virus detection is possible towards translational diagnostics. These spectroscopic features are likely a reflection of the unit structure and molecular content of each virus and can be used to differentiate various types of viruses.

## Conclusion

We have reported a counter-propagating interferometric mid-infrared photothermal imaging system and its application to label-free spectroscopic imaging of a single virus. We harnessed the principle of interferometric scattering to probe the weak signal from a viral particle. Using this approach, we have recorded the fingerprint IR spectra from single poxvirus and vesicular stomatitis virus. The dominant peaks are contributed by the amide I and amide II groups and the peak ratio allows differentiation of the two groups of viruses. Unlike SERS, our approach does not need plasmonic enhancement on a nanostructure substrate. Collectively, interferometric mid-infrared photothermal microscopy as a pump-probe technique opens opportunities for label-free spectroscopic detection of a broad size range of individual viruses and biological nanoparticles.

## Acknowledgments

This work is supported by R01 GM126409, R35 GM136223, and R42 CA224844 to JXC.